\documentclass[12pt,letterpaper]{article}
\usepackage{amsmath,amssymb,pgf,pgfarrows,pgfnodes,float,appendix, hyperref}
\usepackage{graphicx}
\usepackage{subfigure}

\usepackage[margin=0.9in]{geometry}
\usepackage{pgfplots}
\usepackage{amsmath}
\usepackage{tikz}

\newcommand{\beq}{\begin{equation}}
\newcommand{\eeq}{\end{equation}}
\newcommand{\abs}[1]{|#1|}

\title{{\rm\footnotesize \qquad \qquad \qquad \qquad \qquad \ \qquad \qquad \qquad \ \ \ \ \ \                  RUNHETC-2019-13 }\vskip.5in    On the Low Density Regime of Homogeneous Electron Gas }
\author{Tom Banks and Bingnan Zhang\\
Department of Physics and NHETC\\
Rutgers University, Piscataway, NJ 08854\\
E-mail: \href{mailto:bz173@scarletmail.rutgers.edu}{bz173@scarletmail.rutgers.edu}}

\date{}
\begin{document}
\maketitle

\begin{abstract}
We investigate the low density limit of the Homogeneous Electron system, often called the {\it Strictly Correlated} regime.  We begin with a systematic presentation of the expansion around infinite $r_S$, based on the first quantized treatments suggested in the existing literature.  We show that the expansion is asymptotic in the parameter $r_S^{1/4}$ and that the leading order result contains exponential corrections that are significant even for $r_S \sim 100$.  Thus, the systematic expansion is of limited utility.
As a byproduct of this analysis, we find that there is no Wigner Crystal (WC) in one spatial dimension.  This is an example of the Mermin-Wagner theorem, but was not appreciated in some earlier literature.  More modern work\cite{newrefs} has come to conclusions identical to ours.  Note that the long range Coulomb potential modifies the dispersion relation of phonons in one dimension, but still leads to the instability of the crystal, due to a very weak infrared divergence.  We then propose a new approximation scheme based on renormalization group ideas.  We show that the Wegner-Houghton-Wilson-Polchinski exact renormalization group equation reduces, in the low density limit, to a classical equation for scale dependent electron and plasmon fields.
In principle, this should allow us to lower the wave number cutoff of the model to a point where Wigner's intuitive argument for dominance of the classical Coulomb forces becomes rigorously correct.

\end{abstract}

\section{Introduction}

The system of non-relativistic electrons with a uniform positive background charge density, interacting only via Coulomb interactions, is the basic object of Density Functional Theory.  If we use the Rydberg energy and the Bohr radius to introduce dimensionless time/energy and space/momentum coordinates, the first quantized Hamiltonian for this system is 
\beq H = \sum_i \frac{{\bf P}_i^2}{2} + \int d^d x d^d y\  :(N({\bf x}) - n_0) (N({\bf y}) - n_0):\frac{1}{\abs{{\bf x} - {\bf y} }} , \eeq where $ N({\bf x}) = \sum_i \delta^3 ({\bf x - X_i}) $, 
\beq \int d^d x\ (N({\bf x}) - n_0) = 0 , \eeq and the normal ordering symbol means that we leave out the self interaction.  
If we rescale all the coordinates by $a$ then the first term scales like $a^{-2}$, while the second scales like $a^{-1}$ .   The absolute scale of the coordinates is set by the total volume $V$.  We'll use periodic boundary conditions, though everything we say could be done for any other choice.  Thus, the large $a$ limit is equivalent to a large $V$ limit.
The other parameter in this system is the conserved electron number $K = n_0 V$, and the large $V$ limit is taken with $n_0$ fixed.  One defines $n_0 = \frac{A_d}{r_s^d}$, where $A_d$ is the area of the unit sphere in $d$ space dimensions. Consequently, the large $a$ limit, where the Coulomb interactions dominate the electron kinetic energy, is equivalent to the limit of large $r_S$.   The electrons also have spin degrees of freedom, which do not appear in the Hamiltonian.  It is believed that, as a function of $n_0$, the system develops spin polarized phases in the thermodynamic limit.  For the purposes of this paper, we will ignore the spin.  Henceforth, our electrons are spinless.

Wigner argued that at large $r_S$, the electrons would form a crystal, whose properties are calculated by minimizing the classical Coulomb energy of point electrons.  The lattice spacing is of order $r_S$, and the preferred minimum energy crystal is BCC in $d = 3$, triangular or hexagonal in $d = 2$ and equal spaced in $d = 1$.  Of course, the electron kinetic energy is singular in these delta function wave functions.  The textbook treatment of this issue is to mimic what is done in the Born-Oppenheimer approximation: expand the multi-body potential around the crystalline minimum, and solve the resulting coupled oscillator problem exactly. The scaling arguments then suggest that the width of electron wave functions scales like $r_S^{3/4}$, which means that in the large $r_S$ limit, the electrons are localized on scales much smaller than the lattice spacing.  

Our interest in the low density limit was sparked by the realization that the large $N$ approximation (with $N$ the number of electron spin components) misses the WC phase entirely\cite{bz1}.  The reason for this is that in the conventional limit where $N$ goes to infinity, the coupling goes to zero, and everything else is held fixed, the operator $N(x)$ has a continuous spectrum.  Thus, the WC exists only at densities of order $1/N$, outside the range of validity of naive large $N$ methods.  The Hartree approximation becomes exact at large $N$, because anti-symmetrization can be performed solely on the spin components of the wave function.   The exchange corrections captured by the Hartree-Fock approximation do not resolve this problem.  Indeed, if the scaling predictions are correct, then, as we will see, the exchange terms should be exponentially small at large $r_S$.  In fact, the HF approximation predicts a ground state that violates translation invariance, but is not a crystal.  In the HF approximation, the width of electron wave functions at large $r_S$ scales like the putative lattice spacing.

We therefore searched the literature for, but have not yet found, a systematic presentation of the large $r_S$ expansion, based on the idea of expanding the multi-electron potential about its WC minimum.  We were particularly worried about the fact that oscillator wave functions have widths that scale like the inverse square root of the frequency.   In the large volume limit, the long wavelength oscillations have frequencies that scale like $V^{-1/d}$ since these are the phonon modes of spontaneously broken translation invariance.  If the localized density expectation value gets significant contributions from these long wavelength modes, then the electrons are not truly localized on the lattice, and the entire approximation scheme is inconsistent.  We note that Quantum Monte Carlo schemes that work with only a finite number of electrons, are apt to miss the effects of long wavelength phonons, because the effective volume of the QMC system is not large in $r_S$ units.  

In the next section, we will outline the systematic large $r_S$ expansion and find that it is self consistent, though of limited validity, in $d = 2,3$, but that the one dimensional WC phase does not exist.  We show that the large $r_S$ expansion is only asymptotic, with even the leading order result having corrections of order $e^{ - C r_S^{1/2}}$ where the constant $C$ involves a complicated sum over phonon modes of the crystal.
Furthermore, the corrections to the leading order result give wave functions with complicated multi-body correlations, and Fermi statistics adds to the inherent complication of the expansion around the minimum.  Thus, even for $r_S \sim 100$ one must be skeptical of the accuracy of the leading order result, and one is unlikely to be able to systematize the corrections to it.

We've therefore begun searching for a more robust calculational scheme in the WC regime, which might be able to capture the phase transition that ends this regime. Following \cite{kivspiv} we speculated in \cite{bz1} that the transition leads to a colloidal phase.   The essential "flaw" in Wigner's classic argument for the WC is a confusion between infrared physics, dominated by the long range Coulomb potential, and ultraviolet physics dominated by the electron kinetic energy.  In quantum field theory this sort of issue is usually simplified by use of the renormalization group.  We therefore implement the Wegner-Houghton-Wilson-Polchinski (WHWP) renormalization group equation for the homogeneous electron system.  The WHWP equation is an exact equation for the scale variation of the non-quadratic part of the effective action for the quantum fields.  Starting from a simple action $\phi \bar{\psi} \psi$ at a spatial cutoff scale $\delta \ll 1$ in Bohr units it tells us how to find an action with cutoff length $e^l \delta$, which will have the same correlation functions as the original action for all wavenumbers $< e^{-l} \delta^{-1}$.  Expressed in terms of Feynman diagrams, the WHWP equation has only tree and one loop contributions.  We will argue that for $ e^l \delta \ll r_S$ the one loop terms are negligible.  The tree approximation to the WHWP equations can be solved in terms of a scale dependent field configuration satisfying a "classical equation of motion".  

If we consider the original bare action with a cutoff $e^l \delta$, then at the wave function that minimizes the Coulomb term, the expectation value of the kinetic energy is
of order $(e^l \delta)^{-2} $ while the Coulomb energy is of order $1/r_S$.  Thus we need $r_S < e^{2l} \delta^2 $ in order to treat the kinetic energy as a perturbation, while our approximate RG calculation requires $e^l \delta \ll r_S$ .   The consistency of the two approximations requires only that $e^l \delta \gg 1$, which of course implies that $r_S$ is large.  We will show that the higher order interactions induced by the RG calculation are also smaller than the bare Coulomb term for a range of $l$.   Thus, the combination of an approximate solution of the WHWP equation combined with perturbation theory around the Coulomb interaction in the cutoff theory gives us a systematic tool for calculating the properties of the WC.

\section{Large $r_S$ Expansion}

Wigner's argument treats electrons as classical particles.  Indeed, the most straightforward quantum mechanical interpretation of the argument is that the scaling equations tell us that the first quantized Hamiltonian has the parametric dependence on $r_S$ that one associates with semi-classical physics.  The ground state is determined by expanding the variables around a classical solution.  The first quantized Hamiltonian for $K$ electrons has an $S_K$ symmetry.  Fermi statistics is the statement that this is a gauge symmetry:  the only allowed states must lie in a particular one dimensional representation of the permutation group.  

The classical minima of the multi-body Coulomb potential are NOT invariant under $S_K$.  The minimum positions sit on a crystal lattice.  At any particular minimum of the potential, a particular electron sits on a particular lattice site.  There are $K!$ minima of the multibody potential, differing by the choice of which electron sits at which site.  Fermi statistics tells us that the actual state of the system is the anti-symmetric superposition of the ground states for each of these minima.  If, in the true quantum ground state of the distinguishable electron problem, the probability of finding the electron closest to lattice point $i$ is concentrated within a distance of $i$ that is much less than the lattice spacing, then the overlap between different permutations of the electrons is exponentially small.

On the other hand, for a fixed choice of minimum of the multi-electron potential, the ground state wave function is
\beq \psi ({\bf x_1 \ldots x_K} ) = \prod_J \sqrt{1/2\pi \omega_J (q)} e^{ - \omega_J (q_J)^2 } , \eeq , where the $q_J$ are the normal modes of oscillation and $\omega_J$ is the frequency of the $J$th normal mode. There is always a zero mode, corresponding to a slow uniform translation of the whole lattice. The corresponding coordinate is the center of mass $K^{-1} \sum {\bf X}_i$.  In volume $V$ the time-scale for motion of this coordinate goes like $K^{1/2}$ so a Gaussian wave packet fixed at any value will not spread on the scale of internal motions of the non-zero modes.  We can simply freeze it at a fixed value. 

However, as $V \rightarrow \infty$ there are modes of frequency $V^{- 1/d}$ whose wave functions have a width of order $r_S^{3/4} V^{1/d}$  .   For $V^{1/d} > r_S^{1/4}$, which is always true in the thermodynamic limit, the long wavelength modes have widths much larger than the lattice spacing of the WC.   The width of the marginal density distribution
\beq \rho (x) = \int d^{d(K - 1)} y\ \abs{\psi}^2 (x, y_1 \ldots y_{K - 1})\eeq will be dominated by that of the low frequency modes, unless the probability of a low $q$ mode to be concentrated at a particular point goes to zero with $q$.  For 1d, $\omega (q) \sim \abs{q}\sqrt{\abs{\log{q}}}$, the width square of the density distribution will behave like
\beq \int \frac{d q}{\abs{q}\sqrt{\abs{\log{q}}}} \eeq  This integral is divergent, so the density distribution does not have a crystalline structure.  N. Andrei\cite{andrei} pointed out to us that this result should be viewed as an example of the Mermin-Wagner theorem that there is no long range order in one dimension.  It is destroyed by quantum fluctuations of the Goldstone excitations, here the phonons.  This remains true even though the Goldstone dispersion relation is modified by the long range Coulomb interaction, although the divergence is weakened to a square root of the logarithm of the volume.  This would make it hard to see in a QMC calculation with a modest number of electrons.

For $d \geq 2$, the width of the density distribution is smaller by a factor of $r_s^{- 1/4}$ than the lattice spacing, so the WC phase indeed exists.   So far, our discussion of it has neglected Fermi statistics.  To get the true ground state wave function, one must sum over the oscillator wave functions for each classical minimum, with an exchange of which electron sits at which lattice site, multiplying by a minus sign whenever the permutation from the original electron configuration is odd.  Corrections to expectation values of permutation symmetric operators will come from integrals of products of wave functions differing by a permutation.  For $d \geq 2$ these overlaps have the form
\beq \int \psi^* \psi_{perm} = \sum e^{- C_i r_s^{1/2}} . \eeq  The computation of the coefficients $C_i$ involves complicated sums over collective coordinates.   

Higher order corrections to the harmonic approximation are {\it extremely} complicated.
There are an infinite number of anharmonic corrections to the Coulomb potential, and more and more of them must be taken into account in each order of the $r_S$ expansion around a given minimum.  It is highly unlikely that even the series around a given minimum is convergent, because the asymptotic behaviors of the harmonic and Coulomb potentials are infinitely different at large separation.  In the fermionic problem, the contributions to the ground state energy coming from overlaps between the wave functions at different minima clearly have essential singularities at $r_S = \infty$.   The expansion around the classical Wigner crystal, which gives the large $r_S$ asymptotics of the ground state energy of the HEG is thus both extremely complicated to compute\footnote{It is unclear to us whether the leading approximation, including exponentially small corrections, has ever been computed exactly.  The literature we have found\cite{ldlit} does not give enough detail to determine whether the formulae quoted are exact or approximate.} and merely asymptotic.  Furthermore, the parameter $r_S^{-1/4}$, which controls the size of corrections to the leading order, is not small even for $r_S \sim 100$ where the melting transition of the WC is claimed to occur based on Quantum Monte Carlo (QMC) results.  This means that the expansion is unlikely to be a useful guide to physics near that transition.  

\subsection{Harmonic oscillator analysis of low density HEG}
Here we record the computations whose results we have discussed in the previous subsection. 
\subsubsection{1d}
\ \ \ \  For simplicity, consider 2N+1 electrons on a line with length L, and identify the two ends of the line. There is also uniform positive background charge distributed on the line to make the energy extensive. However, since the background potential is everywhere constant, we can neglect the background  in the following analysis.
\par The Lagrangian is  
\begin{equation}
\mathcal{L}=\sum_i\frac{\dot{x_i}^2}{2}-\frac{1}{2}\sum_{i\neq j}\frac{1}{|x_i-x_j|}
\end{equation}
where $i$ runs over all electrons. Notice that the distance is defined in a periodic system, so $|x_i-x_j|$ is actually $|x_i-x_j\  \mod\  L|$, and it must be no larger than $\frac{L}{2}$.
\par Suppose the electrons have a small deviation $y_i$ from their equilibrium position $L_i$, we expand the Lagrangian to second order
\begin{equation}
\mathcal{L}=\sum_i \frac{\dot{y_i}^2}{2}-\frac{1}{2}\sum_{i\neq j}\frac{1}{|L_{ij}+y_{i}-y_j|}\approx \sum_i \frac{\dot{y_i}^2}{2}-\frac{1}{2}\sum_{i\neq j}\frac{1}{|L_{ij}|^3}(y_i-y_j)^2
\end{equation}
where $L_{ij}=|L_i-L_j|$ is the distance between i and j when $y_i=y_j=0$.
Fourier transform it to momentum space $y_j=\sqrt{\frac{1}{2N+1}}\sum_k \beta_k e^{i kj}$, where $k=-\frac{2\pi N}{2N+1},-\frac{2\pi(N-1)}{2N+1},...,\frac{2\pi (N-1)}{2N+1},\frac{2\pi N}{2N+1}$
\begin{equation}
\begin{aligned}
\mathcal{L}=&\sum_k \frac{1}{2}\dot{\beta_k}^2-\frac{1}{2(2N+1)}\sum_{kq}\sum_{j}\sum_{m=-N,m\neq 0}^N\frac{1}{|L_m|^3}(2\beta_k\beta_{-q}-2\beta_k\beta_{-q}e^{ikm})e^{ij(k-q)}\\
=&\sum_k\frac{1}{2}\dot{\beta_k}^2-\frac{1}{2}\sum_{k}\sum_{m\neq 0,m=-N}^N \frac{1}{|L_m|^3}2(1-e^{ikm})\beta_k\beta_{-k}
\end{aligned}
\end{equation}
$A_k\dot{=}\sum_{m\neq 0,m=-N}^N \frac{2(1-e^{ikm})}{|L_m|^3}=\sum_{m\neq 0,m=-N}^N \frac{2(1-\cos{km})}{|L_m|^3}$ is real. $|L_m|=|mr_S|$.
$\beta_k$ is complex,$\beta_k^*=\beta_{-k}$ . Write $\beta_k=(a_k+ib_k)/\sqrt{2}$.
Choose a frame such that the center of mass doesn't move, the zero mode is $\beta_0=0$.
\begin{equation}
\mathcal{L}=\frac{1}{2}\sum_{k>0}(\dot{a_k}^2+\dot{b_k}^2)-\frac{1}{2}\sum_{k>0}A_k (a_k^2+b_k^2)
\end{equation}
\par After quantization, this is a collection of 2N independent harmonic oscillators. The ground state wave function is 
\begin{equation}
\phi=\prod_{k>0}(\frac{1}{2\pi \sigma_k^2})^{1/2}e^{-\frac{a_k^2+b_k^2}{4\sigma_k^2}}
\end{equation}
where $\sigma_k^2=\sqrt{\frac{1}{4A_k}}$. And the probability distribution is Gaussian. Since $a_k, b_k$ are independent random variables, their linear combinations  also obey Gaussian distributions. From $y_j=\sqrt{\frac{2}{2N+1}}\sum_{k>0}[a_k\cos(kj)-b_k\sin(kj)]$, we know \begin{equation}
\sigma(y_j)^2=\frac{2}{2N+1}\sum_{k>0}\sigma_k^2=\frac{1}{(2N+1)}\sum_{k>0}\frac{1}{\sqrt{A_k}}
\end{equation}
For small momentum modes $0<k<k_s\ll 1$, the dispersion can be simplified
\begin{equation}
\omega_k^2=A_k=\sum_{m=1}^N \frac{4(1-\cos{km})}{(mr_S)^3}\approx\sum_{m=1}^{f/k}\frac{2k^2}{mr_S^3}\approx -\frac{2}{r_S^3}k^2\log{k}
\end{equation}
where $0<f<1$ is a cutoff. Terms with $m>f/k$ are suppressed by $1/m^3$, and  can be neglected.
\begin{equation}
\sigma(y_j)^2=\frac{1}{(2N+1)}\sum_{k=\frac{2\pi}{2N+1}}^{\frac{2\pi N}{2N+1}}\frac{1}{\sqrt{A_k}}>\frac{1}{(2N+1)}\sum_{k=\frac{2\pi}{2N+1}}^{k_s}\frac{1}{\sqrt{A_k}}\approx\int_{\frac{2\pi}{2N+1}}^{k_s}dk\ \frac{1}{2\pi k}\sqrt{\frac{r_S^3}{-2\log{k}}}\approx\sqrt{\frac{r_S^3\log{N}}{2\pi^2}}
\label{eq:1d1}
\end{equation}

So if we fix the density, and send the electron number to infinity, the  packet width of single electron probability distribution will diverge. The lattice configuration breaks down before the thermodynamic limit is achieved. 
\par We didn't antisymmetrize the wave function in the above calculation. Under the crystal assumption, $\sigma(y_i)$ scales as $r_S^{\frac{3}{4}}$, so the overlap between different permutations is exponentially suppressed at large lattice spacing. The conclusion (\ref{eq:1d1}) doesn't change even if we consider antisymmetrization.
\subsubsection{Higher dimensions}
\ \ \ \ Generalization to 2d is straightforward. In this case the preferred WC is a triangular lattice. Pick two directions as coordinate axes, and suppose the lattice size is $(2N+1)(2M+1)$. The Fourier transformation reads $\vec{y}_{\vec{j}}=\frac{1}{\sqrt{(2N+1)(2M+1)}}\sum_{\vec{k}}\textbf{C}_{\vec{k}}\cdot\vec{\beta}_{\vec{k}}e^{i\vec{k}\cdot \vec{j}}$, where $\vec{k}=(k_1,k_2)$ is momentum, $k_1=-\frac{2\pi N}{2N+1},-\frac{2\pi (N-1)}{2N+1},...,\frac{2\pi N}{2N+1}\ k_2=-\frac{2\pi M}{2M+1},...,\frac{2\pi M}{2M+1}$, $\textbf{C}_{\vec{k}}$ is a $2\times 2$ matrix. When N and M are large, we can replace $\frac{1}{(2N+1)(2M+1)}\sum_{\vec{k}}=\int\frac{d^2 \vec{k}}{4\pi^2}$.Using the same method as in the one dimensional case, one can show that the single electron distribution function is also Gaussian, and the square of its packet width is
\begin{equation}
\sigma^2 \sim\int d^2\vec{k} \frac{1}{\sqrt{\sum_{\vec{m}\neq 0}\frac{1-cos(\vec{k}\cdot \vec{m})}{(|r_S\vec{m}|^3)}}}<\int d^2\vec{k}\frac{1}{\sqrt{\frac{2-cos(k_1)-cos(k_2)}{r_S^3}}}\approx \int_{small\  k}d^2 \vec{k} \frac{\sqrt{2}r_S^{3/2}}{|\vec{k}|}
\end{equation}
which is finite. In 3d it is $\sigma^2< \int d^3 \vec{k} \frac{r_S^{3/2}}{|\vec{k}|}$, also finite. So there is no divergence in higher dimensions.  This implies that the WC exists above one dimension, for sufficiently large $r_S$. 

As we've noticed, the actual wave function is the antisymmetrized sum of products of harmonic oscillator ground states for the normal mode coordinates of the lattice, with different electrons assigned to different sites.  If we look at an individual electron coordinate, its wave function is concentrated around one lattice point for one assignment and another one for a different assignment.  Thus, overlaps between different wave functions in the anti-symmetrized sum will give a series of terms of the form $A_i e^{ - c_i r_S^{1/2}}$.  The computation of the coefficients $A_i$ and $c_i$ is quite arduous and we have not performed it.  Thus, even the leading large $r_S$ contribution to the ground state energy is not analytic around $r_S = \infty$ and the expansion is only asymptotic.

Computation of the next order term in the expansion of the ground state energy requires one to compute the expectation value of the general fourth order symmetric polynomial in electron coordinates in the complicated leading order sum of Gaussian wave functions that we have just described. Higher orders require expectation values of higher order symmetric polynomials, as well as matrix elements of symmetric polynomials between different energy levels of the normal mode oscillators, again in anti-symmetrized sums.  In the thermodynamic limit we also have to worry about the contributions in higher orders of phonon modes nearly degenerate with the zeroth order ground state.  Even if these don't give rise to infrared divergences, they can easily make nominally higher order terms in the expansion comparable in size to lower order terms, and are apt to introduce dependences on ${\rm ln}\ r_S$ in the expansion.

In summary, the large $r_S$ expansion establishes the existence of the WC phase in two and three dimensions, but it is an extremely messy asymptotic expansion whose nominal expansion parameter is $r_S^{-1/4}$.  That parameter is $\sim 1/3$ even for $r_S \sim 100$.  It seems very unlikely that this expansion will be of utility for studying the phase structure of the model at finite values of $r_S$.

\subsection{Quantum Monte Carlo Calculations}

The most widely accepted description of the physics of the HEG in the strongly coupled low density regime is based on some form of QMC calculation\cite{QMC}.  There are a number of well known issues with this method.  It requires that one works separately in regions of coordinate space where the wave function has a fixed sign, where the boundaries of those regions are determined by an initial Jastrow wave function from which the Monte Carlo relaxation begins.  Phase transitions are determined by comparing the variational energies of wave functions with different symmetry properties.
QMC methods reproduce the physics of the weakly coupled high density phase in a satisfactory manner.  At strong coupling they have no clear competitor so it's hard to determine their accuracy.  

Our concern about the reliability of QMC methods stems from two sources.  The first has to do with the fact that, the total linear size of the system one is able to study, in Wigner lattice units, is rather small.  Thus, the method is likely to miss physical effects associated with long wavelength phonons, or the long wavelength fluctuations near any second order phase transition.  Secondly, in \cite{bz1} we provided evidence for the claim of \cite{kivspiv} that the HEG has a colloidal phase.  The analysis of \cite{bz1} used the bubble nucleation description of first order phase transitions, as well as notions of continuum elasticity theory, applicable to collective behavior of large numbers of electrons. It is not at all clear how one would model such a phase in a QMC calculation with a small number of electrons.  Furthermore, the colloidal picture provided an intuitive picture for the gapless excitation at non-zero wave number that has been found in two different methods of resumming Feynman diagrams\cite{ht} , at a density far above that at which the WC disappears in the QMC calculations.  We argued that the colloid makes a smooth transition from a gel regime, consisting of a crystal with bubbles of fluid trapped inside it, to a sol regime, consisting of chunks of crystal immersed in fluid.  We showed that if the surface tension of the crystal chunks was negative, as expected from the repulsion of the surface electrons and screening of charge in the bulk of the crystal, that these chunks were stable and had lower energy than an equivalent volume of fluid.

Using this picture, we argued that the second order transition occurs roughly at a place where the negative surface tension vanishes.  The size of the chunks goes to zero in this limit and their classical energy goes to zero.  We conjectured that these semi-classical crystalline excitations become gapless Bose quasi-particles in the zero tension limit.  The transition from the fluid phase to the sol-colloid is Bose condensation of these particles and the fact that the gapless excitation occurs at non-zero wave number is the residual signature of the fact that these excitations can be viewed as classical chunks of crystal away from the phase transition.  QMC calculations could not reveal such a phase and the associated transition, without an ansatz for the initial wave function that incorporates colloidal physics.  We do not know how to construct such an ansatz, and it would seem to involve much larger numbers of electrons than one can deal with, given current computational resources.   For example, if there is a sol phase at a value of $r_S$ where the typical number of electrons in a crystalline chunk is of order the number of electrons in the QMC simulation, then it's possible that QMC would mistake this regime for a crystalline phase.

One piece of evidence for the existence of a phase of the HEG which is neither a crystal, nor a translation invariant fluid, is the negative compressibility found in QMC calculations\cite{QMC}\cite{book}\cite{negcompress} at densities far above that at which the WC melts.  Negative compressibility is accompanied by a negative static dielectric function, which is impossible in a translationally invariant phase\cite{negdielnot}.  In fact, the Feynman diagram resummations of\cite{ht} all show that the negative compressibility/dielectric function sets in at precisely the point at which a gapless mode with non-vanishing wave number occurs.  QMC calculations seem to be a very good guide to the physics at very high and very low densities, and it's plausible that they remain reliable at densities of order the melting transition point.  The negative compressibility found by these calculations, at densities somewhat above the melting point suggests the existence of a non-crystalline phase, which is not invariant under translations, in agreement with the diagram resummation methods.

\section{Renormalization Group Analysis} 
\subsection{Introduction to the RG Analysis}

In our opinion, what is missing in the analysis of the HEG is a systematic treatment of the strongly coupled low and medium density regime of the system.  The straightforward low density expansion, which we reviewed in the first section of this paper, is extremely complicated and non-convergent.  In addition, internal evidence from the leading term leads one to expect that it is unreliable even in the vicinity of the melting transition.  In this section, we will propose a new method, which is completely unbiased by symmetry patterns of variational ansatze, and is in principle amenable to systematic improvements.   The basic problem with Wigner's intuitive argument for the dominance of the Coulomb interaction is that the expectation value of the kinetic term in the Hamiltonian is infinite in the eigenstates of the Coulomb piece of the Hamiltonian, and so is manifestly not a small perturbation.  The systematic low density expansion resolves this problem by the classic method of Born and Oppenheimer - expansion of the many body potential around its minimum.  Instead we will view the problem as a classic example of the necessity of disentangling ultra-violet and infrared physics.  This is the problem for which the Renormalization Group was invented.

Our basic idea is very simple.  We will argue that at sufficiently low density, the physics of the short wavelength modes of the electron and plasmon fields becomes soluble, so that one can "integrate out these modes" and obtain an action with a relatively long ultraviolet wavelength cutoff, much longer than the Bohr radius, though much shorter than $r_S$.  With this cutoff, the kinetic term really is a small perturbation of the Coulomb interaction.   The method thus combines two approximations; an approximate solution of the RG equations determining the effective action for the long wavelength modes, followed by a perturbative determination of the long wavelength ground state, around the WC.  Obviously, the approximations can only be reliable if $r_S$ is large enough that the system is in a phase where the WC is at least metastable.  They also involve a choice of the cutoff length scale $1 \ll L \ll r_S$ (in Bohr units), and the reliability of finite order truncations might depend on the choice of $L$, or on the precise form of the cutoff at the microscopic scale $l$.  

The intuitive reason that the RG equations are easy to solve at low density is that in the UV physics, only the short range part of the Coulomb potential is taken into account, but electrons are far apart from each other.  Thus, the short range physics has interactions that can be treated perturbatively at low density.  In order for this intuition to work we must introduce the cutoff in a way that remains fairly local in position space, so that a sharp wavenumber cutoff is inappropriate.  We will discuss a non-relativistic version of the Wegner-Houghton-Wilson-Polchinski\cite{WHWP}(WHWP) non-perturbative RG equations, which lead to transparent analytical formulae.  In the conclusions we will suggest that a Kadanoff style lattice block spin approach might be more amenable to numerical calculations.  

Before proceeding to the technical discussion of the RG, we should mention another popular approach to the strongly coupled regime.  This is the Strictly Correlated Electron problem, which gives an {\it upper} bound on the density functional or the Born-Oppenheimer potential.   The bound on the DF is the sum of the minimum values of the kinetic and Coulomb terms in the Hamiltonian, at fixed expectation value of the density operator $N(x)$.  Since it takes the Coulomb energy into account exactly, one expects this upper bound to be close to the true value for $r_S$ in the WC regime.  Note that it does not suffer from UV divergences because the expectation value of the kinetic term is part of the density functional that one tries to minimize.  We will review this approximation and its suitability for finding a colloidal phase in an Appendix .  

\subsection{The WHWP Equation for the HEG}

The derivation of the WHWP equation starts by replacing the photon and fermion propagators in Feynman diagrams with the following substitutions
\beq \frac{1}{p^2} \rightarrow \frac{1}{p^2}f(p^2 \delta^2) , \eeq
\beq \frac{1}{i\omega - k^2 / 2 + \mu} \rightarrow  \frac{1}{i\omega - \delta^{-2}c(\delta)} F(k^2 \delta^2) . \eeq   We are working in imaginary time.  The wavelength cutoff $\delta $ is $ \ll 1$ .   The coefficient $c$ is tuned as a function of $\delta$ so that the energy per electron is a finite function of $\mu$, the chemical potential\footnote{Alternatively we could normal order the interaction term in the Hamiltonian.}.  The smooth functions $f$ and $F$ approach $1$ at small wavenumbers and go to zero faster than any power of wavenumber at large values of their argument.
The choice of these functions changes the detailed solution of the WHWP equations, but not their qualitative nature.  In high energy physics this is an example of a choice of {\it renormalization scheme}.  It's possible that numerical implementation of our procedure might be more accurate for a particular choice of scheme, but we are not yet able to say anything useful about this point.  We will see that the analysis of the equations is simplest if we choose the first derivative of both functions to vanish.

Starting from the microscopic interaction $S_I (\delta) = i\int d^4 x\ \phi (x) \psi^{\dagger} (x) \psi (x)$\footnote{For notational convenience we relabel the imaginary time variable as $x^0$.} we ask the question of whether we can rescale the cutoff in the propagator to $e^s \delta$, and change $S_I$ in such a way that correlation functions of the fields at wavenumbers $< e^{-s} \delta^{-1}$ are unchanged\footnote{Actually, for the kind of cutoff function we're using, these functions are changed by amounts of order $e^{- k^2 \delta^2}$.   To leave them completely unchanged we must make the propagator a smooth function that  vanishes identically above the cutoff momentum.}.   

In the functional integral formalism, correlation functions of the fields in imaginary time are calculated as 
the expansion coefficients of
\beq Z[J] = \int [d\Phi\ d\psi\ d\psi^{\dagger}]\ e^{ - [S_{\delta} + S_I + \int \ [\eta^* \psi + \eta \psi^{\dagger} + J \Phi]] } . \eeq  We omit the normalizing factor that sets $Z[0] = 1$ because it doesn't contribute to the connected correlation functions, which are the expansion coefficients of $W = {\rm ln}\ Z$.  $S_{\delta} $ is the quadratic action that gives rise to the cutoff propagators above.  We want to replace this with $S_{e^{s} \delta}$ and compensate for that by replacing  $S_I = i \Phi \psi^{\dagger} \psi$ by an $s$ dependent interaction, as long as we restrict the source functions $\eta, \eta^* $ and $J$ to have support only for wave numbers below $e^{-s} \delta^{-1}$.

We can write the equation for the partial derivative of the connected generating functional $W$ as

\beq \partial_s W = Z^{-1} \int [dF_i] [\frac{e^{S_I}}{2} \int d^4 p\ F_i (p) F_j (- p) \partial_s D_{ij} (p) + \partial_s (e^{S_I} ) ] e^{\frac{1}{2} \int d^4 p F_i (p) F_j (-p) D_{ij} (p) + F_i (p) J^i ( - p)} . \eeq  $Z$ is the generating functional with sources set to zero. In this equation, for compactness of notation, we have introduced the three component field $F_i \equiv (\Phi, \psi, \bar{\psi})$.  Although some of its components are Grassmann fields, all the manipulations that we do will involve pairs of such fields and so all the potential minus signs cancel out.  The inverse propagator $D_{ij} (p)$ only has plasmon-plasmon and fermion-(anti) fermion components.  The WHWP observation is that if we take
\beq  \partial_s (e^{S_I} ) = \frac{1}{2} \int d^4 p\  \partial_s D^{-1}_{ij}  (p) \frac{\delta^2}{\delta F_i (p) \delta F_j (-p)} (e^{S_I} ) , \eeq then we can integrate by parts in the functional integral and rewrite the derivative of the interaction in terms of the action of this second order differential operator on $e^{\frac{1}{2} \int d^4 p F_i (p) F_j (-p) D_{ij} (p) + F_i (p) J^i ( - p)}$.  This action gives
\beq \begin{aligned} e^{\frac{1}{2} \int d^4 p\ F_i (p) F_j (-p) D_{ij} (p) + F_i (p) J^i ( - p)}\ &\frac{1}{2} \int d^4 p\  [\partial_s D^{-1}_{ij}  (p)D_{ij} (p) + F_i (p) D_{ij} (p) \partial_s (D^{-1}_{jk} (p)) D_{kl} (p) F_l (- p)\\& + J_i (p) J_j (-p) \partial_s D^{-1}_{ij} (p) ] .\end{aligned}  \eeq

The first term gives a source independent contribution to $W$, which is a contribution to the ground state energy of the system.  Using the identity $\partial_s D^{-1} = - D^{-1} \partial_s D  D^{-1}$ we see that the second term cancels the term coming from the $s$ dependence of the kinetic term in the action.  Finally, the third term contributes only to the connected two point function and the contribution is concentrated up near the cutoff $e^{-[s + {\rm ln}\ \delta]} $. 

For those more comfortable with Feynman diagrams than functional integrals, the same equation can be derived diagrammatically.  The $s$ derivative acts sequentially on each propagator in a diagram.  If the propagator connects two parts of a one particle reducible diagram then the result sums up to the term quadratic in $S_I$ in the WHWP equation.  Otherwise it's in a closed loop and gives a contribution to the second term.
The reason that only tree and one loop diagrams contribute to the scaling derivative of the action is that the $s$ derivative of a propagator carrying loop momentum is non-zero only at the cutoff momentum and above, while the propagator itself vanishes rapidly above the cutoff, so the region of integration is restricted, up to exponentially small corrections, to a small region around the cutoff.  At low density, there is a further simplification.  Any diagram containing a fermion in a closed loop has an energy integral, which restricts the electron momentum to be less than the fermi momentum.  
Mathematically this is due to the fact that we can close the energy contour in the complex plane, and the poles of the integrand all lie below the fermi momentum. Since the fermi momentum is well below the cutoff scale, whereas the internal momenta are all at the cutoff scale, these diagrams are proportional to the cutoff derivatives of propagators, evaluated at the fermi momentum.  They are therefore exponentially small as long as the cutoff $e^s \delta \ll r_S$, since the latter scale determines the size of the fermi momentum. There is a single exception to this rule.  The one loop diagram with a single electron line and a single plasmon line, the electron self energy, has an energy integral that does not converge rapidly enough to close the integration contour in the complex plane.  This diagram leads to renormalization of the chemical potential, and we tune the coefficient $c$ in the electron propagator (that is, make it $s$ dependent) in such a way that the expectation value of the density is equal to the physical value $\frac{4\pi}{3 r_S^3}$.  All other loop corrections vanish at low density.

We can therefore re-write the WHWP RG equation at low density as
\beq \partial_s S_I = \int d^4p [\frac{\delta S_{I}}{\delta \psi (p)} \partial_s G (p)  \frac{\delta S_{I}}{\delta \bar{\psi} (-p)} +  \frac{\delta S_{I}}{\delta \Phi (p)} \partial_s D(p)  \frac{\delta S_{I}}{\delta \Phi (- p)} .\eeq  This is the functional analog of a first order non-linear partial differential equation, and as such it may be solved by the method of characteristics.   That is, there are scale dependent field configurations $\bar{\psi} (p,s), 
{\psi} (p,s), {\Phi} (p,s)$ such that
\beq S_I [ \bar{\psi} (p), 
{\psi} (p), {\Phi} (p) ; s] = S_I [\bar{\psi} (p,s), 
{\psi} (p,s), {\Phi} (p,s) ; 0] . \eeq   To see this, note that the scale derivative of the right hand side is
\beq \begin{aligned}\partial_s S_I [\bar{\psi} (p,s), 
{\psi} (p,s), {\Phi} (p,s) ; 0] =& \int d^4 p\ (\partial_s\bar{\psi} (p,s)\frac{\delta}{\delta \bar{\psi (p,s)}} +
\partial_s {\psi} (p,s)\frac{\delta}{\delta {\psi (p,s)}}+ \partial_s {\Phi} (p,s) \frac{\delta}{\delta \Phi (p,s)})\\& S_I [\bar{\psi} (p,s), {\psi} (p,s), {\Phi} (p,s) ; 0] .\end{aligned} \eeq
This is the WHWP equation if
 \beq \partial_s\bar{\psi} (p,s) = \partial_s G(p) \frac{\delta}{\delta \psi (p,s)} S_I [\bar{\psi} (p,s), {\psi} (p,s), {\Phi} (p,s) ; 0] , \eeq
\beq \partial_s \psi (p,s) = \partial_s G(p) \frac{\delta}{\delta \bar{\psi} (p,s)} S_I [\bar{\psi} (p,s), {\psi} (p,s), {\Phi} (p,s) ; 0] , \eeq
\beq \partial_s \Phi (p,s) = \partial_s D(p) \frac{\delta}{\delta \Phi (p,s)} S_I [\bar{\psi} (p,s), {\psi} (p,s), {\Phi} (p,s) ; 0] . \eeq   Solving these equations by the usual method of iterated integration, we get a sequence of corrections to the bare interaction $S_I = i\int d^4 x\ \Phi (x) \bar{\psi (x) \psi (x)}$ when written in terms of the fields at $s = 0$.    It's easy to verify that the higher order terms become small if $s$ is large, except for a term proportional to the first derivative of $f$ or $F$ at zero wave number. 

To see this, we recall that the bare interaction is
\beq S_I = i\int d^4 p\ d^4 q\ \Phi (q) \bar{\psi} (p)\psi (p + q) . \eeq
The characteristics of the approximate WHWP equation have the form
\beq \partial_s F_i (s,p)  = \partial_s P_{ij} (s,p) C_{ijk} \int d^4 q\ F_j (s,q) F_k (s,p + q) .\eeq  $F_i$ are the three independent fields, $P_{ij} $ the propagator connecting $F_i$ to $F_j$ and $C_{ijk} = 1$ if all the indices are different and vanishes otherwise.  This is equivalent to the set of integral equations
\beq F_i (s,p)  = F_i (0,p) + \int_0^s dr\ \partial_r P_{ij} (r,p) C_{ijk} \int d^4 q\ F_j (r,q) F_k (r,p + q) .\eeq These can be solved by iteration.   If either of the functions $f$ or $F$ has a non-vanishing first derivative at the origin, then there is a term on the RHS of the WHWP differential equation, which does not vanish for $p^2 \delta^2 = 0$.  That is $\partial_s P_{ij} (s,p) \neq 0$ at $p = 0$.  Keeping only this term, we can Fourier transform the equation to position space
\beq F_i (s,x)  = F_i (0,x) + \int_0^s dr\ \partial_r P_{ij} (r,0) C_{ijk} \ F_j (r,x) F_k (r,x) .\eeq This equation is {\it ultralocal} in position space.  The new fields at a point are ordinary functions of the old fields.  This means that the correlation functions of $F_i (s,x)$ at points separated by distances large compared to the rescaled cutoff $e^s \delta$ are dominated by those of $F_i (s,0)$ .  The long distance behavior, which determines the phase structure and transport properties of the system is completely unaffected by these renormalizations.

This is consistent with the fact that, if we choose the first derivatives of the cutoff functions to vanish, there are no ultra-local contributions at all.  Assuming this is the case, it is easy to see that the iterative solution to the integral form of the WHWP equations generates a series of interactions proportional to higher and higher powers of $(e^s \delta p)$, which is what high energy physicists call a systematic effective field theory expansion.   

For the application to the physics of the low density electron fluid, and thereby to the Born Oppenheimer approximation, we want to start with the bare interaction at a scale $\delta$ below {\it e.g.} the Bohr radius of high $Z$ atoms and choose the final value of $s$ such that $e^s \delta \ll r_S$ .  According to QMC calculations, the density above which the Wigner crystal is not the ground state appears to be at around $r_S = 50$ in two dimensions and $r_S =100$ in three dimensions.  Thus we probably need the maximal value of $s$, $s_{max}$ to be around $8-9$.  The wave numbers in the theory with cutoff $\Delta$ are of order $e^{- s_{max}} \delta^{-1}$ or smaller so the higher order terms in the iterative solution of the WHWP equations are smaller than the term that gives the bare Coulomb interaction.

To summarize, we can approximate the WHWP equations for the HEG by a simpler equation, up to exponentially small corrections $\sim e^{- c\frac{r_S^2}{e^{2s_{max}} \delta^2}}$, whose details depend on our choice of the cutoff function at scale $\delta$.  For applications to atomic, molecular and condensed matter physics, it seems reasonable to take $\delta \sim 10^{-3}$ .   That is, we write the HEG Hamiltonian with a length scale cutoff of one thousandth of a Bohr radius. Solving the simplified equations, we get a sequence of corrections to the simple two body Coulomb interaction, which are suppressed by powers of $e^{-s_{max}}$.  

Our proposal for the solution of the homogeneous electron model at low density is then to show that, with cutoff $e^{s_{max}} \delta$, the effective action for the plasmon field $\Phi$, obtained by integrating out the electrons, has a periodic solution, corresponding to a crystal, for large enough $r_S$.   The classical plasmon field will not be delta function localized because of our Gaussian momentum cutoffs.  Fluctuations of $\Phi$ around the classical solution will describe the phonon modes of the crystal.   The fermion wave functions in this background will have expectation values of the kinetic energy per particle  of order $[e^s \delta]^{-2}$ while the periodicity of the solution is $r_S$ and the Coulomb energy per particle of order $r_S^{-1}$.  Thus, we can expect this configuration to have lower energy than a homogeneous one when $r_S < e^{2s} \delta^2$.  This is compatible with the inequality $r_S \gg  e^s \delta $, which guarantees that the density profile has a distinct lattice structure, when $e^s \delta \gg 1$.   Remarkably, this crude estimate of the transition density is roughly the value given by QMC calculations.  

In the extreme low density limit we've seen that the width of the peaks in the density distribution scale like $r_S^{3/4}$, with a numerical coefficient that involves the coupling of a single electron coordinate to all of the normal mode frequencies.   Thus, if we choose $e^s \delta \sim r_S^{3/4}$ our RG procedure should be able to reproduce the physics of this extreme limit.   

\section{Conclusions}

In this paper we have set up two approximation schemes for studying the low density regime of the HEG.  The first is the systematic large $r_S$ expansion, in which one expands the $K$ electron Coulomb interaction around its $K!$ WC minima, and solves the resulting coupled oscillator problem for each minimum as the zeroth order approximation.  The correct wave function is the antisymmetric superposition of the normal mode ground states for these minima.   This resembles, but is not the same as the Jastrow ansatz. There are a number of important results that follow from this analysis.

\begin{itemize}

\item The expansion parameter is $r_S^{-1/4}$, which is not small near the transitions out of the WC state found by QMC calculations.  Furthermore, even the zeroth order approximation to the ground state energy has corrections of the form $A_i e^{- c_i r_S^{1/2}}$, where the constants $c_i$ and the prefactor of the exponential are hard to calculate. The expansion is thus asymptotic and complicated and the expansion parameter is not small even for $r_S \sim 100$.

\item  In one dimension there is no Wigner crystal.  The low frequency normal mode wave functions have widths larger than the lattice spacing and in one dimension the density expectation value is dominated by these contributions, washing out the crystal structure.  This is an example of the Mermin-Wagner theorem about one dimensional long range order.

\item The higher order corrections to the leading order calculation are extremely difficult to calculate.  They lead to complicated many body correlations .

\end{itemize}

These considerations make it unlikely that the large $r_S$ expansion will be a useful
tool for studying the possibility of the colloidal phases discussed in \cite{bz1}, or indeed any of the phase structure of the HEG.  In two and three dimensions it establishes the existence of the WC phase at large $r_S$, and that may be the limit of its utility.

Instead, we proposed a scheme for studying the entire low density regime using the tools of the Wegner-Wilson-Polchinski exact renormalization group equations.  We argued that these equations simplify in the large $r_S$ regime.  Starting from a length scale cutoff $\delta \ll 1$ in Bohr units, we argued that there is a systematic expansion of the effective Hamiltonian for the HEG in powers and exponentials of $e^{s}$, where $e^s \delta$ is a cutoff scale $< r_S$ but large enough that all terms in the Hamiltonian are small perturbations of the cutoff bare Coulomb interaction.  The detailed form of the corrections depends on the choice of the smooth cutoff function introduced in the WHWP scheme.  Thus there is a complicated but systematic perturbation theory, entirely in the second quantized formalism (unlike the large $r_S$ expansion).   We suspect that this expansion might be convergent.

We will reserve a detailed study of this approximation scheme to future work, but we want to mention an interesting idea connected with it, which might speed up convergence of the expansion.   The WHWP formalism introduces an arbitrary smooth cutoff function $f$.  When using the equations to find fixed points of the RG flow, it is well known that the fixed points and their spectra of dimensions are independent of the choice.   That will not be true in the present case, where we are only using the RG to integrate out a finite range of length scales.   This suggests that using the effective Hamiltonian of the RG as a variational ansatz might choose a form for $f$ that optimizes convergence to the exact answer.   A particularly appealing form of this idea is to reformulate the RG equations with a lattice cutoff and to use the Tensor Network Renormalization Group. We hope to return to this problem.

\section{Appendix: Strictly Correlated Electrons}

In the Born-Oppenheimer approximation, we are interested in finding the quantum state that minimizes the expectation value of the operator
\beq H = K + C + \int N(x) V(x) , \eeq where $N(x) = \psi^{\dagger} (x) \psi (x)$ is the electron density operator, $K$ is the kinetic energy and $C$ is the Coulomb repulsion between the electrons.   The minimum is the B-O potential. Write $H$ as
\beq [K + \int N(x) U(x)] + [C + \int N(x) (V(x) - U(x))] \equiv  H_1 + H_2, \eeq where $H_{1,2}$ are the two operators in square brackets.  For any choice of $U(x)$ the 
B-O potential is {\it greater} than the sum of the individual minima of $H_{1,2}$.  It's been argued \cite{SCE} that the bound is close to being saturated at large $r_S$, for a choice of $U(x)$ that is determined by requiring that the two different minimizing states give the same density expectation value.  It's easy to see that that criterion is the same as asking for the largest lower bound among those obtained in this manner.  The resulting estimate of the B-O potential is called the Strictly Correlated Electron or SCE approximation.

This way of formulating the SCE approximation has several computational advantages, which we just want to sketch in this appendix.  Minimization of $H_1$ is simply the problem of solving the Schrodinger equation in an external potential.   This can be done analytically in both the small and large $U$ limits (the latter limit is controlled by the JWKB approximation as long as the potential does not vary too wildly).  Minimization of $H_2$ is best done by putting the system on a lattice with spacing much smaller than the Bohr radius.  It is then equivalent to finding the ground state of the classical Ising model in the presence of a spatially varying source.  This can be done efficiently in both the limits $V - U \gg 1$ and $V - U \ll 1$, and is of course a completely classical physics problem.  Quantum mechanics enters it only through the discrete values of the lattice density operator.  

The limit of small $V - U$ should be useful for studying single atoms.  There the minimization of $H_1$ is the exactly soluble Bohr atom if we take $U$ to be equal to $V$ and reduces to a collection of one dimensional Schrodinger equations if $U$ is a general spherically symmetric potential.  An interesting choice for $U$ might be the self consistent Hartree-Fock potential, which minimizes the full atomic Hamiltonian, restricted to the subspace of Slater determinant wave functions.  $V - U$ might then be smaller than the electron self interaction everywhere, allowing us to use a perturbative evaluation of the expectation value.

More generally, if $V(x)$ is the nuclear potential of a system with a number of high $Z$ nuclei, we can take $U(x)$ to be the potential of nuclei of large charge.  The minima will clearly be the Bohr atomic ground states of those atoms (neglecting electron electron repulsion), up to corrections exponential in the distance between the highly charged nuclei.  The length scale in the exponentials can be calculated from textbook single particle wave functions.   

Finally, we'd like to note that the SCE approximation, since it is an approximation to the density functional, is well suited to discussing the hypothetical colloidal phases of Kivelson and Spivak.

\begin{center} 
{\bf Acknowledgments }  \end{center}

This work is partially supported by the U.S. Department of Energy and 
 TB would like to thank N.Andrei,E.Rabinovici, S.Sachdev, D. Vanderbilt, E. Lieb, P. Gori Giorgi, K. Burke,\  M.Fisher,S.Kivelson, B.Spivak, and especially K.Haule for crucial insights and helpful suggestions, and the entire condensed matter group at Rutgers university for patiently listening to an interloper on their territory and making numerous helpful comments.

\end{document}